\documentclass[usenatbib]{mn2e}
\usepackage{color} 
\usepackage{graphicx}
\usepackage{epsfig}
\usepackage{natbib}

\begin{document}

\title[Galaxies with unusually young and massive stars] 
{A study of 1000 galaxies with unusually young and massive stars
in the SDSS: a search for hidden black holes}   

\author [G.Kauffmann] {Guinevere Kauffmann$^1$\thanks{E-mail: gamk@mpa-garching.mpg.de}
, Claudia Maraston$^2$, Johan Comparat$^3$, 
Paul Crowther$^4$\\
$^1$Max-Planck Institut f\"{u}r Astrophysik, 85741 Garching, Germany\\
$^2$Institute of Cosmology, University of Portsmouth, Burnaby Road, Portsmouth PO1 3FX, UK\\
$^3$Max-Planck-Institut für Extraterrestrische Physik, Giessenbachstrasse 1, 85748, Garching, Germany\\
$^4$Department of Physics and Astronomy, University of Sheffield, Sheffield S3 7RH, UK\\}

\maketitle

\begin{abstract} 
We select 1076  galaxies
with extinction-corrected H$\alpha$ equivalent widths too large to   
be explained with a Kroupa (2001) IMF, and compare these with a
control sample of galaxies that is  matched in stellar mass, redshift and
4000 \AA\ break strength, but with normal H$\alpha$ equivalent widths. Our
goal is to study how
processes such as black hole growth and energetic feedback processes from
massive stars differ between galaxies with extreme central
H$\alpha$ emission and galaxies with normal young central stellar
populations. The stellar mass distribution of H$\alpha$ excess galaxies is
peaked at $3 \times 10^{10} M_{\odot}$ and almost all fall well within
the star-forming locus in the [OIII]/H$\beta$ versus [NII]/H$\alpha$ BPT disgram. 
H$\alpha$ excess galaxies are twice as likely to exhibit
H$\alpha$ line asymmetries and 1.55 times  more likely to be detected
at 1 GHz in the VLA FIRST survey compared to control sample galaxies. 
The radio luminosity per unit stellar mass decreases with the stellar age of the system.  
Using stacked spectra,  we demonstrate that [NeV] emission is  not present in the very youngest of 
the radio-quiet H$\alpha$ excess galaxies with detectable Wolf-Rayet features, suggesting that black hole growth has not yet commenced in such systems. 
[NeV] emission is detected in H$\alpha$ excess galaxies with radio detections
and the strength of the line correlates with the radio luminosity. 
This is the clearest indication for a population of  black holes that may be forming
in a subset of the H$\alpha$ excess population.  
\end {abstract}

\section{Introduction}

The majority of accreting black holes identified through their UV/optical
emission appear to be accreting at sub-Eddington rates, both in the local
Universe (Kauffmann \& Heckman 2009) and at redshifts $z$=1-2 (Aird et al 2011).
The existence of very luminous quasars at $z \sim 6-7$ that are likely
powered by accretion onto black holes of masses of at least $10^9 M_{\odot}$
(Fan et al 2003) implies a separate channel of ``seed'' black hole formation in
the early Universe (Volonteri 2010) to allow black holes to reach very high
masses less than a gigayear after the Big Bang.

The physics governing the formation of seed black holes is still highly
speculative.  Many theoretical models have been presented in the literature with
varying assumptions and complexity. The models generally fall into two
categories: 1) direct collapse of low angular momentum gas clouds (e.g.  Loeb \& Rasio
(1994), Bromm \& Loeb (2003),  Begelman, Volonteri \& Rees (2006), Lodato \&
Natarajan (2006,), 2) mergers and accretion in dense stellar clusters
(e.g.  Portegies Zwart et al 2004, Devecchi \& Volonteri 2009).

Most observational work has focused on detecting a {\em remnant} population of
seed black holes of masses $10^4 - 10^6 M_{\odot}$ (often termed imtermediate
mass black holes -- IMBHs for short) in present-day galaxies (see Mezcua 2017
for a detailed review). There have been observational campaigns focused on the
detection of IMBHs through stellar kinematic studies of globular clusters, or
via signatures of gas accretion at X-ray wavelengths
in nearby galaxies. 

Much recent
activity has also been devoted to identifying subsets of dwarf galaxies that are likely
to harbour low mass black holes. Out of a sample of 44,594 galaxies with stellar
masses less than $3 \times 10^9 M_{\odot}$ in the Sloan Digital Sky Survey,
Reines et al (2013) identify a subsample of 151 galaxies with emission line
spectra indicative of ionization by an AGN (i.e 0.3 \% of the parent sample).
Twenty-five of these objects exhibit broad emission lines, and the width of these
lines, together with the assumption that the ionized gas is in virial
equilibrium within the potential well of the black hole, provide upper limits on
the black hole masses of $\sim 10^5 M_{\odot}$.  The detection of hard X-ray
emission spatially coincident with core radio emission in a subset of these
dwarf galaxies (Reines et al. 2011, 2016; Reines \& Deller 2012; Baldassare et al.
2015) provides futher strong evidence for the existence of accreting black holes
in these systems.  

Observational studies that focus on identifying IMBHs in the local galaxy
population do not necessarily shed light on the processes by which they formed.
The majority of present-day galaxies are relatively quiescent and processes such
as gas accretion or runaway stellar mergers that formed the IMBHs may have
occurred quite far in the past. We note however that 
follow-up observations of one of the brightest known
ultra-luminous X-ray sources ESO 243-49 HLX-1 in the SOa galaxy
ESO 243-49 revealed the presence of a surrounding
young stellar population of age $\sim$13 Myr (Farrell et al 2012).  

Another possible route to studying the formation of IMBHs is to search
for evidence of accreting black holes in young star clusters.
Motivated by 
evidence that the stars near the Milky Way's
center have  IMFs with an excess of massive stars (e.g. Lu et al 2013,
Hosek et al 2019),     
Kauffmann (2021) carried out a search for galaxies
in the range $10^{10}-10^{11} M_{\odot}$ with
central stellar populations indicative of an initial mass function (IMF)
flatter than Salpeter at high stellar masses.  
15 face-on galaxies with stellar masses in
the range $10^{10}-10^{11} M_{\odot}$ were identified in the 
2nd public data release of the Mapping Nearby Galaxies at APO (MaNGA) survey
(Bundy et al 2015) where  the 4000 \AA\ break was  either flat or rising towards
the centre of the galaxy, indicating that the central regions host evolved
stars, but where the H$\alpha$  equivalent width was also steeply rising to extremely
high values in the central regions.  The ionization parameter was 
low in these unusual Galactic Centres, indicating that ionizing sources were primarily
stellar rather than AGN.  Wolf-Rayet features characteristic of hot young stars
were also often found in the spectra and these tended to get progressively
stronger at smaller galactocentric radii. Finally, a large fraction of
these objects (8 out of 15)  were detected at radio wavelengths. 

There are a number of other suggestions from observations  that the stellar
initial mass function (IMF) in environments with high stellar densities may
differ from the canonical Kroupa (2001) IMF. Globular clusters in the Andromeda
galaxy exhibit a trend between metallicity and mass-to-light ratio that only a
non-canonical, top-heavy IMF could explain (Haghi et al.  2017) and
ultra-compact dwarf galaxies have large dynamical mass-to-light ratios and
appear to contain an overabundance of luminous X-ray binary sources
(Dabringhausen et al. 2009).  
As pointed out in a recent theoretical paper
by Weatherford et al. (2021), star clusters that have formed with a top heavy IMF
are expected not only to produce more black holes, but also to produce many more
binary black hole mergers and intermediate mass black holes.  Natarajan (2021)
also point out that gas accretion during the initial formation of the cluster
can lead to extremely rapid growth, scaling a stellar mass remnant seed black
hole up to intermediate mass black hole range.  If the star-forming bulge
already contains a pre-existing central supermassive black hole, its mass is
also likely to be growing significantly if surrounded by starburst with a
top-heavy IMF.

Kauffmann (2021) highlighted one galaxy out of the sample of 15 as a possible
candidate for a ``transition" object with a black hole in the process of
formation or rapid new growth. This galaxy is among those with the strongest
Wolf Rayet signatures. It was unusual in that it showed the strongest
Wolf-Rayet signatures, significant central HeII$\lambda$4686 emission and
clear non-Gaussian H$\alpha$ profiles in the centre of the galaxy.

In this paper, we return to the full sample of galaxies with single-fibre
spectroscopy from Sloan Digital Sky Survey observations. We select all galaxies
with extinction-corrected H$\alpha$ equivalent widths well above the range that
can be explained with a Kroupa (2001) IMF, and compare these objects with a
control sample of galaxies that is exactly matched in stellar mass, redshift and
4000 \AA\ break strength, but with ``normal" H$\alpha$ equivalent widths. Our
goal is to use the much larger samples to investigate more systematically how
processes such as black hole growth and energetic feedback processes from
massive stars may differ between galaxies with extremely strong central
H$\alpha$ emission and galaxies with ``normal" young central stellar
populations. We also use stacked spectra to investigate the presence of
``hidden'' AGN in galaxies where the strongest emission lines are dominated by
emission from HII regions.

In section 2, we describe the selection of the two samples and discuss their
distributions of physical properties such as stellar mass, $r$-band light
concentration index, mean stellar age, and their locations in the Baldwin, Philipps \&
Terlevich (1981) (BPT) emission line ratio diagrams. We also  present the
quantities that are measured directly from the galaxy spectra.  In section 3, we
analyze non-Gaussian H$\alpha$ line profiles, blue and red bump Wolf Rayet
features in the two samples, as well as correlations between these quantities
and the global properties of the host galaxies.  A cross-match with the VLA
FIRST catalogue (Condon et al 1998) allows us to analyze the radio properties of the
samples.  In section 4, we analyze stacked spectra constructed from H$\alpha$
excess galaxies with radio and Wolf Rayet feature detections. In section 5, we
summarize our results and discuss the future prospects of this work.

\section {Sample selection and derived quantities} We begin with two
publically-available value-added galaxy catalogues distributed through the SDSS
Science Archive Server (SAS). The first is the Wisconsin catalogue of PCA-Based
Stellar Masses and Velocity Dispersions (Chen et al 2012), which contains 
estimates of star formation history parameters (mean stellar age
and burst mass fraction), metallicity, dust extinction and
velocity dispersion, based on a library of model spectra for which principal
components have been identified. In this paper, we make use of the stellar
masses and mean stellar ages estimated using Bruzual \& Charlot (2003) stellar
population synthesis models with a Kroupa (2001) IMF
to generate the principal components.

The second is the catalogue of Portsmouth Stellar Kinematics and Emission Line
Fluxes (Thomas et al 2013).  
The Penalized PiXel Fitting (pPXF, Cappellari \& Emsellem 2004) and Gas and
Absorption Line Fitting code (GANDALF v1.5; Sarzi et al. 2006) is  used to
calculate stellar kinematics and to derive the emission line properties of
galaxies.  The stellar population models from Maraston \& Str\"omb\"ack (2011)
based on the MILES stellar library (S\'anchez-Bl\'azquez et al. 2006), augmented
with theoretical spectra at wavelengths $<$ 3500 \AA\  from Maraston et al. (2009),
based on the theoretical library UVBLUE (Rodr\'iguez-Merino et al. 2005) are
adopted as templates for stellar continuum fitting. In this paper, we make use
of the H$\alpha$ flux and equivalent width measurements, as well as the strong
line fluxes H$\beta$, [OIII]$\lambda$5007 and [NII]$\lambda$6584 to examine the
location of galaxies in the standard BPT line diagnostic diagram.

We select two galaxy samples on the basis of their H$\alpha$ equivalent widths.
Following the procedure adopted in Kauffmann (2021), when calculating the
H$\alpha$ equivalent width, the H$\alpha$ line flux is corrected for dust
attenuation using the measured Balmer decrement using the formula $A_V=1.9655
R_V \log(H\alpha/H\beta/2.87)$, where $R_V=3.1$ and the Calzetti (2001)
attenuation curve has been adopted. The stellar continuum measurements are not
corrected for dust attenuation. This simplifying assumption is
motivated by the finding by Wild et al (2011) that there is a strong 
increase in emission-line-to-continuum dust attenuation
with the specific star formation rate of the galaxy.  
The samples are limited in redshifts to $z<0.37$
to ensure that H$\alpha$ lies well within the wavelength range where secure
measurements of the H$\alpha$ equivalent width are possible, and to galaxies
with $M_*>10^9 M_{\odot}$.

The first sample is selected to have EQW(H$\alpha$/\AA)$>800$.  As shown
in Kauffmann et al (2201), this is the maximum H$\alpha$ equivalent achievable for 
a single stellar population of age $\sim 10^6$ yr.  This sample is termed the
H$\alpha$ excess sample and consists of 1076 galaxies selected out of a total of
857493 galaxies in the same redshift and stellar mass range (i.e. 0.125\% of the
parent sample). The second sample is selected to have $80<$EQW(H$\alpha$/\AA)$<300$ 
and consists of 10710 galaxies. In Figure 1, red and blue  histograms show how
this samples and the H$\alpha$ excess  sample are  distributed as a function of a
variety of different quantities. These include 1)stellar mass $M_*$, 2)redshift,
3)4000 \AA\ break strength (we use the narrow definition in Balogh et al (1999),
which we term D$_n$(4000)), 4)mean stellar age, 5)concentration index of the
$r$-band light, defined as the ratio of the aperture enclosing 90\% of the total
light to the aperture enclosing 50\% of the light, 6)half-light radius
of the galaxy in arcsecond, 7)Balmer
decrement H$\alpha$/H$\beta$, and 8)the two BPT line ratios [OIII]/H$\beta$ and
[NII]/H$\alpha$. 

The H$\alpha$ excess sample shown by red histograms is clearly offset to higher stellar
masses, redshifts and Balmer decrement values compared to the  sample 
with normal H$\alpha$ EQW values, shown by blue histograms.
Interestingly, the stellar mass distribution of the H$\alpha$ excess sample
peaks at $\log (M_*/M_{\odot}) \sim 10.5$, which is the  mass where the galaxy population
transitions from predominantly star-forming to predominantly passive, with
little ongoing star formation (Kauffmann et al 2003a). Only 25\% of the
H$\alpha$ excess sample have stellar masses less than $10^{10} M_{\odot}$ compared
to 60\% of the sample with lower H$\alpha$ equivalent widths. This shows that
the H$\alpha$ excess phenomenon is not confined to low mass galaxies.
The median redshift of the galaxies in the H$\alpha$ excess sample is
$\sim 0.2$ and the galaxies typically have half-light radii less than 2 arcseconds.
The SDSS fiber diameter is 1.5". In very low redshift galaxies, it will sample
light mainly from the inner bulge, but this is not the case for the galaxies
under study here. We note that the median redshift of the sample studied
by Kauffmann (2021) was 0.05. We will come back to the implications of the
mismatch in physical scales for the interpretation of the results in this
paper in the final discussion section of this paper.

\begin{figure}
\includegraphics[width=92mm]{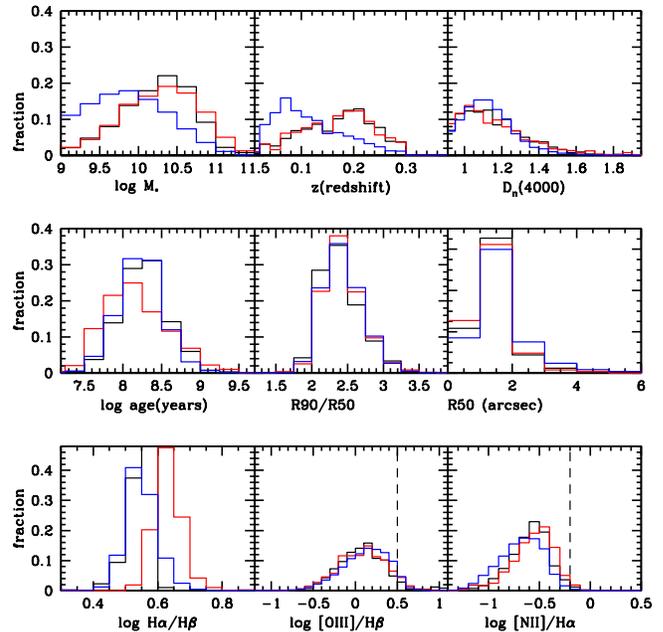}
\caption{Red and black histograms show how
the H$\alpha$ excess and control samples are  distributed as a function of
stellar mass $M_*$, redshift, 4000 \AA\ break strength, mean stellar age, 
$r$-band concentration index, half-light radius in arcseconds, Balmer decrement
and the two BPT line ratios [OIII]/H$\beta$ and
[NII]/H$\alpha$. Blue histograms show the distribution for the full sample with  
80$<$ EQW(H$\alpha$/\AA)$<$300 (i.e. before matching by stellar mass, redshift and 
D$_n$(4000))
\label{models}}
\end{figure}

\begin{table*}
\caption{Table of catalog quantities for the H$\alpha$ excess sample.
The columns are as
follows: 1) SDSS identifier, Plate ID, 2) SDSS identifier, MJD, 3) SDSS indentifier, fibre ID,
4) right ascension (J2000), 5) declination (J2000), 6) redshift,
7) logarithm of the stellar mass (M$_{\odot}$), 8) 4000 \AA\  break (narrow definition),
9) logarithm of the stellar age (years), 10) H$\alpha$ equivalent width,
11) H$\alpha$  equivalent width (extinction corrected), 12) log H$\alpha$/H$\beta$,
13) log [OIII]/H$\beta$, 14)log [NII]/H$\alpha$}
\resizebox{\textwidth}{!}{%
\begin{tabular}{r|c|c|c|c|c|c|c|c|c|c|c|c|r}
\hline			
plate id & mjd & fibre id & RA & DEC & z & log M$_*$ & D$_n$(4000) & log age & EQW(H$\alpha$) & EQW(H$\alpha_c$) 
& log H$\alpha$/H$\beta$ & log [OIII]/H$\beta$ & log [NII]/H$\alpha$ \\ \hline
  266 &51602&521&146.386& 1.147& 0.270&10.246& 1.080& 8.614&  227.77& 1249.76& 3.941&-9.000&-0.832\\
  267 &51608&493&148.146& 0.249& 0.082&10.400& 1.371& 8.618&  140.73& 1223.52& 4.299& 0.105&-0.535\\
  270 &51909&224&152.409&-0.160& 0.138&10.059& 1.005& 7.885&  150.51&  816.37& 3.933& 0.273&-0.604\\
\hline  
\end{tabular}}
\end{table*}

We have also created a {\em matched} control sample by selecting galaxies from the
lower EQW(H$\alpha$) sample so that the two samples match in stellar mass,
redshift and 4000 \AA\ break strength. This sample is shown as a black histogram
in Figure 1.  
As can be seen, a larger fraction of the H$\alpha$ excess sample has mean stellar
ages less than $10^8$ years compared to the control sample. The H$\alpha$ excess sample
also has higher Balmer decrements, but has similar  
morphological parameters and BPT line ratio distributions compared
to the control sample. Figure 2 shows the
location of the galaxies in the H$\alpha$ excess and control samples in the
two-dimensional [OIII]/H$\beta$ versus [NII]/H$\alpha$ BPT diagrams. As can be
seen, almost all the galaxies in both samples fall below the Kauffmann et al
(2003b) demarcation between star-forming galaxies and AGN/composite systems and
none fall in the region of the diagram occupied by Seyferts or LINERs. The
excitation of the strong emission lines in both samples appears to be radiation
from young stars. Figure 3 shows the location of both samples in the plane of
mean stellar age versus stellar mass. Mean stellar age increases with stellar
mass in both samples.  Galaxies with stellar masses of $\sim 10^9 M_{\odot}$
have mean stellar ages of a few $\times 10^7 M_{\odot}$, increasing to a few
$\times 10^8 M_{\odot}$ at a stellar mass of $\sim 10^{11} M_{\odot}$. The
relation is somewhat steeper for the H$\alpha$ excess sample, but the
differences between the relations for the two samples are quite small,
suggesting that the two samples may be indicative of different phases of a
single type of starburst. Note that all references to the control sample from now
on, refer to the sample that is matched in  stellar mass,    
redshift and 4000 \AA\ break strength.
Table 1 provides a list of catalogue quantities for all galaxies
in the H$\alpha$ excess sample. A corresponding table for the control sample
galaxies is also provided as part of the online supplementary material
that accompanies this paper. 

\begin{figure}
\includegraphics[width=92mm]{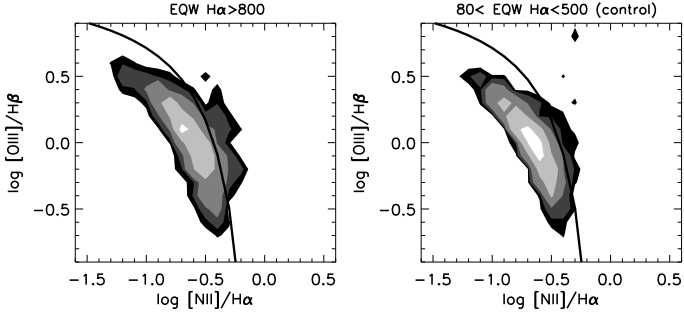}
\caption{Location of the H$\alpha$ excess sample (left) and control
sample (right) in the  [OIII]/H$\beta$ versus [NII]/H$\alpha$ BPT diagrams.
The solid curve shows the Kauffmann et al
(2003b) demarcation between star-forming galaxies and AGN/composite systems.
\label{models}}
\end{figure}

\begin{figure}
\includegraphics[width=92mm]{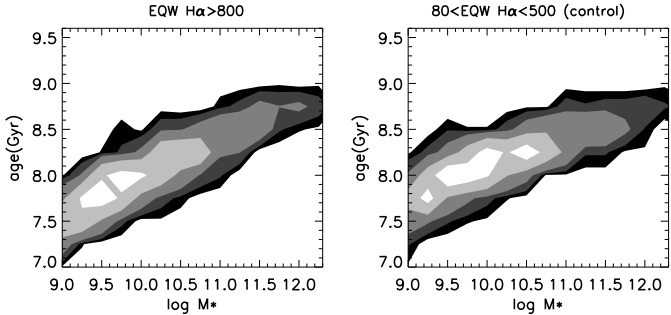}
\caption{Location of the H$\alpha$ excess sample (left) and control
sample (right) in the plane of
mean stellar age versus stellar mass.  
\label{models}}
\end{figure}

\subsection{H$\alpha$ line profile analysis} By construction, the H$\alpha$ line
in the spectra of the galaxies in both samples is always very strong and hence
amenable to a detailed line profile analysis. Most of the standard SDSS spectral
pipelines fit simple Gaussian profiles to a list of strong emission lines after
subtracting a linear combination of template spectra that provides the best fit
to the stellar continuum over the wavelength interval from $\sim$3500 to
$\sim$7000 \AA. In this analysis, we look at the deviations of the observed
spectrum from the best single Gaussian fit.

We extract the observed and the best-fit model spectra obtained using the
FIREFLY code (Wilkinson et al. 2017; Comparat et al 2017) for the galaxies in
the H$\alpha$ excess and control samples.  The left panels in Figure 4 show
three example spectra plotted over the wavelength range from 6530 to 6600 \AA,
which includes the H$\alpha$ line at 6563 \AA, as well as the [NII] lines at
6584 and 6548 \AA. The right panels show the spectra after subtraction of the
FIREFLY continuum fit.  We first fit the two [NII] lines over the wavelength
range shown as red curves in each panel. The width of the weaker 6548 \AA\ line
is fixed to be the same as that of the 6584 \AA\  line.  The best-fit single
Gaussian profiles are subtracted and we then fit the H$\alpha$ line. The
best single Gaussian fit is shown in cyan in each panel in the right column of
the plot. The results shown in the top right panel indicate that a good fit to
H$\alpha$ is obtained with a single Gaussian.  The fits shown in the middle and
bottom right panels reveal significant residual flux away from the line centre.
In the middle right panel, the residual flux is found redwards of the H$\alpha$
line centroid and in the bottom right panel, residual flux is found on both
sides of the line.

\begin{figure}
\includegraphics[width=92mm]{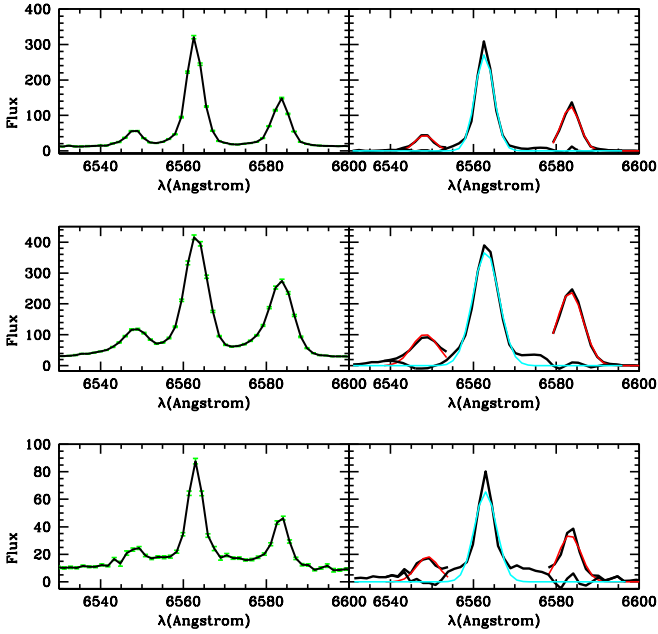}
\caption{Three example spectra plotted over the wavelength range from 6530 to 6600 \AA.
The left panels show the observed spectrum in black with errors indicated as
green errorbars.
In the  right panels, the black curves show the spectra after subtraction of the
FIREFLY continuum fit. The two [NII] lines are fit over the wavelength
range shown as red curves in each panel.The
best single Gaussian fit to the H$\alpha$ line after
subtracting the Gaussian fits to the two [NII] lines is shown in cyan in each panel.
\label{models}}
\end{figure}

We calculate the summed residual flux as the sum of the difference between the
observed flux and the single Gaussian model fit over each spectral bin.  The
residuals are calculated separately on the red and blue side of the line
centroids. If a positive summed residual flux is detected with a $S/N$ greater
than 5 on either the blue or the red side of the H$\alpha$ line centroids, we
classify the galaxy as having an H$\alpha$ asymmetry.  We calculate
$V_{80}({\rm blue})$ and $V_{80}({\rm red})$  in units of km/s by measuring the wavelength
difference between the centroid of the single Gaussian fit to the  H$\alpha$
line  and the wavelength enclosing 80\% of the summed residual flux on both
sides of the line centre.

We note that our procedure differs from that used in other studies (e.g.
F\"orster-Schreiber et al 2019), who fit three broad Gaussian components to
H$\alpha$ and the two [NII] lines.  Their sample includes many classical AGN
with high [NII]/H$\alpha$ ratios, whereas the [NII] lines are always relatively
weak in comparison to H$\alpha$ in our samples. In our sample, it is reasonable
to ascribe the bulk of the detected residual flux to the H$\alpha$ line. We have
also checked our single-Gaussian fits to the [OIII] lines and we almost never
find excess flux at large velocity separations, suggesting that the gas that we
detect at large separation is cool. Finally, as we show in section 4, the high
velocity gas traced by H$\alpha$ is found in conjunction with significant shifts
in the Na I $\lambda\lambda$ 5890, 5896 (Na D) absorption lines with respect to
galaxies that do not exhibit H$\alpha$ line profile asymmteries.  The NaD
absorption lines are believed to trace the neutral gas component of galaxies
(Heckman \& Lehnert 2000).

Finally, we note that the H$\alpha$ line fitting procedure is not  applied to
galaxies where more than 10\% of the spectral bins over the wavelength range
shown in Figure 4 are flagged as problematic. This reduces the size of the
analyzed sample from 1076 to 340 galaxies  located at lower redshifts where
contamination by sky lines are less severe.

\subsection {Identification of Wolf-Rayet Features} The procedure we adopt is
similar to that described in Brinchmann, Kunth \& Durret (2008). 
We adopt the
continuum and central bandpass definitions for the blue and red Wolf-Rayet
features given in Table 1 of this paper. 
The central passband used to probe the blue feature extends over the wavelength range 4655-4755 \AA.
This definition  assumes that the blue Wolf Rayet  feature is dominated by 
broad HeII$\lambda$4686 from WN stars and [FeIII]$\lambda$4658 from O stars and that  the WC stars
that contribute to CIII $\lambda$4650 at the edge of the blue central band are sub-dominant.
The red feature central bandpass assumes that it is dominated by CIV $\lambda$5808 from 
early WC stars, rather than CIII $\lambda$5696 which dominates in late 
WC stars and forms part of the continuum band. 
The definitions are optimized for  metal poor populations where WN stars
dominate over WC stars and where early 
WC stars are relatively common and late WC stars are rare/absent.
We will discuss the limitations of, and possible changes to, this approach 
in the final section of the paper.

Because the galaxy spectra are quite
complex over the wavelength range of the blue and red bump features, 
we rescale the FIREFLY stellar continuum model
fits so that the average flux computed in the two continuum bands in the
observed spectra and in the model fits match exactly. Unlike Brinchmann et al,
we do not attempt to fit indivdual lines within the central bandpass -- instead,
we define an integrated equivalent width for the summed flux over the full
wavelength range. This allows us to calculate the error on the Wolf Rayet
``bump'' detection in a much more straightforward way.  We select all galaxies
with a positive equivalent width measurement with a signal-to-noise greater than
3 as Wolf-Rayet galaxy candidates. Because the features are weak in most of the
galaxies, the analysis in this paper will focus on the analysis of stacked
spectra of subsets of these candidates.

In Figure 5, we plot examples of three blue bump Wolf Rayet candidates from the
H$\alpha$ excess sample  where the central bandpass excess is detected with
$S/N>10$. In the plot, the central bandpass is delineated by blue lines and the
two continuum bandpasses by red lines.  In the bottom panel, we plot the stacked
spectrum of all the blue bump Wolf Rayet candidates.  Only in the stacked
spectrum is the broadened HeII line at 4686 \AA\ clearly visible. There is 
also a weak NIII$\lambda$4640 line visible in the stacked spectrum.  We return to
a discussion of this in section 3.2.  Although our methodology does pick up some
red bump Wolf Rayet candidates (Figure 6), individual emission lines cannot be
distinguished within the red bump window even in the stack. In particular the 
CIV$\lambda$5808 line is not clearly detected even in the stacked spectrum.

\begin{figure}
\includegraphics[width=92mm]{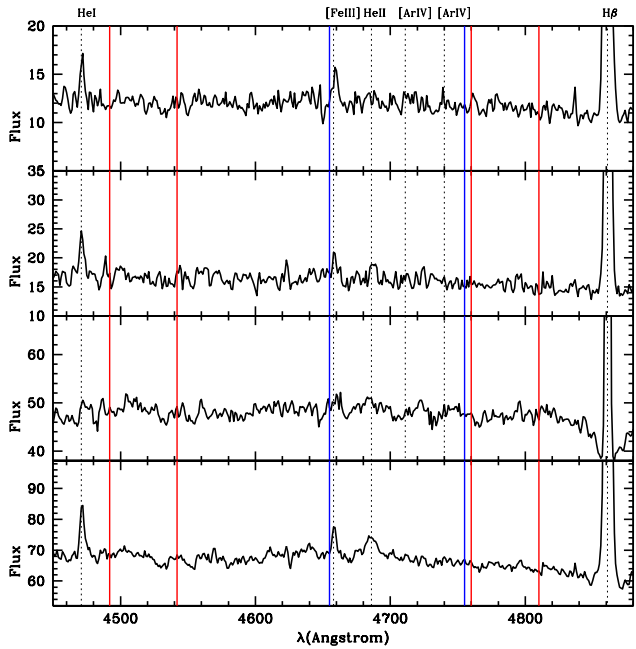}
\caption{The top three panels show examples of three blue bump Wolf Rayet candidates
with $S/N > 10$  from the
H$\alpha$ excess sample. The central bandpass is delineated by blue lines and the
two continuum bandpasses by red lines. The bottom panel shows the stacked
spectrum of all candidates of similar $S/N$   where a broadened HeII line at 4686 \AA\ is clearly visible.   
\label{models}}
\end{figure}

\begin{figure}
\includegraphics[width=92mm]{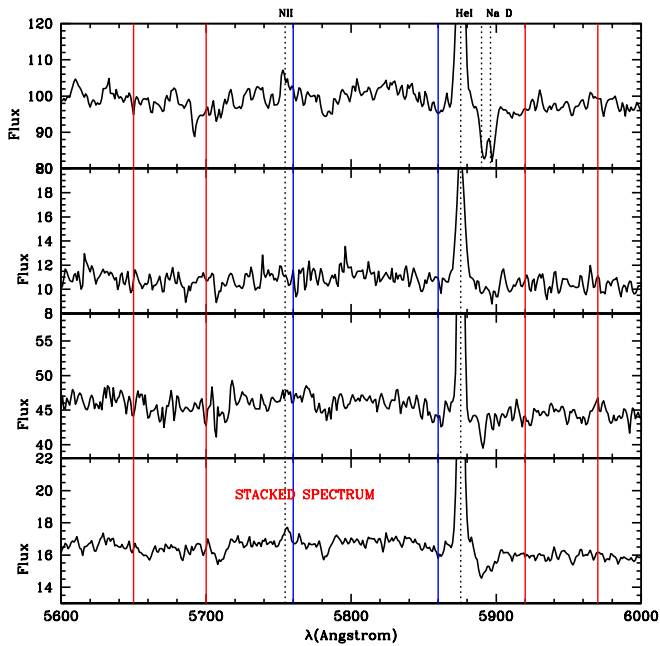}
\caption{The top three panels show examples of red bump Wolf Rayet candidates
with $S/N > 10$  from the
H$\alpha$ excess sample. The central bandpass is delineated by blue lines and the
two continuum bandpasses by red lines. The bottom panel shows the stacked
spectrum of all candidates of similar $S/N$.   
\label{models}}
\end{figure}

In a few  MaNGA galaxies with red Wolf Rayet features studied by Kauffmann (2021),
a narrow CIV$\lambda$5808 emission line was clearly detected in the red bump window and the
the strength of the feature increased strongly towards the central regions
of the galaxy.  The CIV$\lambda$5808 line was narrow
indicating an O star rather than  Wolf Rayet star origin. Nevertheless, an increase in emission from
massive young stars combined with a flat or centrally rising 4000 \AA\ break strength
is difficult to reconcile with a universal IMF, so the study of the line emission
in this wavelength interval is of potential interest in constraining the
relative fractions of the very most massive stars that form in the very central
regions of these galaxies.

These MaNGA galaxies were typically located at
redshifts 0.02-0.05, whereas the galaxies in this sample are at median redshift
$z=0.2$ where the 3 arcsec diameter SDSS fibres enclose stars out to a radius of
5 kpc, which is typically well within the disk. If the CIV$\lambda$5808 emission is
mainly confined to the central stellar populations within galactic
bulges, it is not surprising that it is considerably diluted in the
spectra of more distant galaxies.

\subsection {Radio-loud subset} We have cross-matched the H$\alpha$ excess and
control samples with the source catalog from the Faint Images of the Radio Sky
at Twenty-Centimeters (FIRST) survey carried out at the VLA (Condon et al.
1998), The SDSS and FIRST positions are required to be within 3 arc seconds of
each other (see for example, Best et al 2005). 
We recover 208 VLA FIRST survey cross-matches in the
H$\alpha$ excess sample, compared to 134 in the control sample, i.e. a 55\%
higher detection rate. 

Table 2 provides the main derived quantities analyzed in this paper for the 
H$\alpha$ excess sample.  A corresponding table for the control sample
galaxies is also provided as part of the online supplementary material
that accompanies this paper.

\begin{table*}
\caption{Table of derived quantities for the H$\alpha$ excess sample. The columns are as
follows: 1) SN$_{\rm B}$(H$\alpha$), signal-to-noise of H$\alpha$ asymmetric flux detection on blue side, 
2) F$_{\rm B}$(H$\alpha$), fraction of total blue side H$\alpha$ flux in asymmetric component,  
3) V80$_{\rm B}$(H$\alpha$), wavelength enclosing 80\% of the total blue side asymmetric flux,   
4) SN$_{\rm R}$(H$\alpha$), signal-to-noise of H$\alpha$ asymmetric flux detection on red side, 
5) F$_{\rm R}$(H$\alpha$), fraction of total red side H$\alpha$ flux in asymmetric component,  
6) V80$_{\rm R}$(H$\alpha$), wavelength enclosing 80\% of the total red side asymmetric flux,   
7) logarithm of radio power (Watts/Hz), 8) signal-to-noise of Wolf Wayet blue feature detection,
9) equivalent width of Wolf Rayet blue feature}
\resizebox{\textwidth}{!}{%
\begin{tabular}{r|c|c|c|c|c|c|c|c|c|c|c|c|r}
\hline			
 SN$_{\rm B}$(H$\alpha$)& F$_{\rm B}$(H$\alpha$)&V80$_{\rm B}$(H$\alpha$)& SN$_{\rm R}$(H$\alpha$)&
 F$_{\rm R}$(H$\alpha$)&  V80$_{\rm R}$(H$\alpha$)& log P (Watts/Hz) & SN WR$_B$& EQW WR$_B$ \\ \hline
 0.000 &  0.000 &  0.000 &  0.000 &  0.000&   0.000 &  0.000 &  0.000 &  0.000\\
 0.748 &  0.000 &  0.000 &  4.518 &  0.046&6576.753 &  0.000 &  0.000 &  0.000\\
 0.000 &  0.000 &  0.000 &  0.000 &  0.000&   0.000 &  0.000 &  0.000 &  0.000\\
\hline  
\end{tabular}}
\end{table*}

\section {Results for individual galaxies} In the section we compare the
fraction of galaxies with H$\alpha$ line asymmetries, Wolf Rayet features and
radio detections in the H$\alpha$ excess and control samples.  We also study the
correlations between these quantities with a variety of different host galaxy
properties.

\subsection {H$\alpha$ emission line profile asymmetries} We find that 126 out
of 340 galaxies in the H$\alpha$ excess sample have detectable asymmetries,
compared to 69 out of 340 galaxies in the control sample, i.e. the rate is 80\%
higher in the H$\alpha$ excess sample.

In Figure 7, we present the properties of the narrow H$\alpha$ line emission.
The left panel shows the wavelength of the centroid of the single-Gaussian fit.
This is strongly peaked at 6562.8 \AA\ with only handful of  galaxies showing
centroid line shifts of greater than 0.5 \AA. The right panel shows the
distribution of the Gaussian width of the lines; the majority of galaxies have
$\sigma$ in the range 100-150 km/s. The instrumental resolution of the SDSS
spectrograph is 70 km/s, so these measurements correspond to true gas velocity
dispersions in the range 70-130 km/s. These results show that narrow-line
H$\alpha$ emission is detected in all the galaxies in the sample -- there are no
galaxies where the corrected single component line width is greater than 240
km/s. The fact that we are able to detect and quantify residual emission
components is due to the fact that the lines are extremely strong and are
detected with very high $S/N$.

\begin{figure}
\includegraphics[width=92mm]{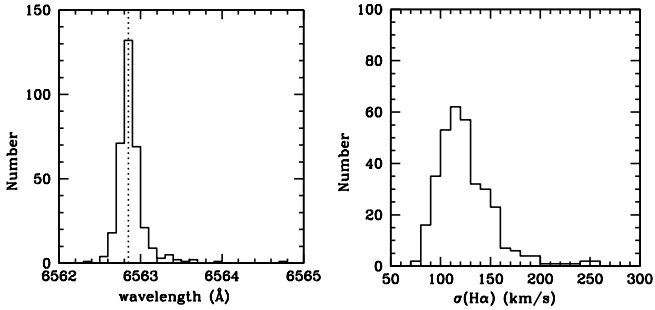}
\caption{The left panel shows the wavelength of the centroid of the single-Gaussian fit.
The dotted line indicates a wavelength of  6562.8 \AA\. 
The right panel shows the
distribution of the Gaussian width $\sigma$ (uncorrected for
instrumental resolution) of the lines.
\label{models}}
\end{figure}

In Figure 8, we plot 1)F(H$\alpha$), the fraction of the total H$\alpha$ flux in
the offset line components and 2)V80, the velocity separation enclosing 80\% of
the asymmetric flux, as a function the stellar mass and the mean stellar age of
the galaxies in the H$\alpha$ excess sample. If excess flux is detected on both the
blue and the red side of the H$\alpha$ line centroid, we plot the maximum of the
two measurements. Later on we will explore whether there are differences between
the H$\alpha$ emission bluewards and redwards of the line centroid.

\begin{figure}
\includegraphics[width=92mm]{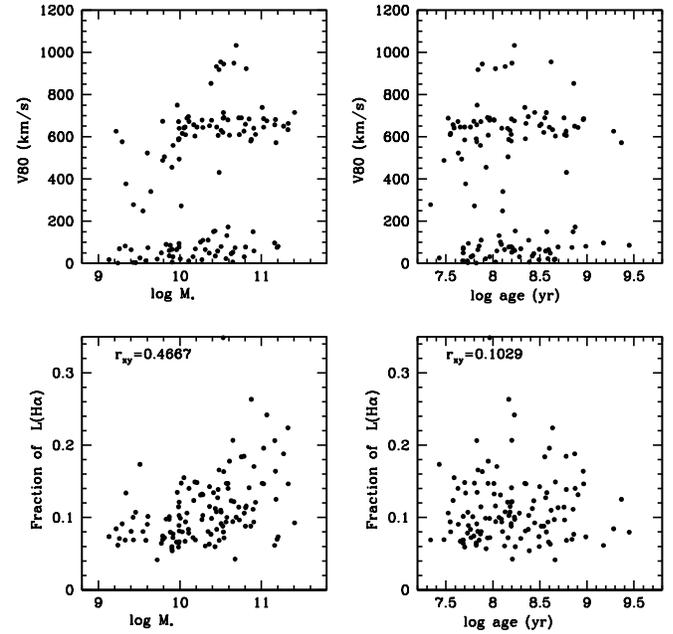}
\caption{The top panels show V80, the velocity separation enclosing 80\% of
the asymmetric flux, as a function the stellar mass and the mean stellar age of
the galaxies in the H$\alpha$ excess sample. 
The bottom panels show F(H$\alpha$), the fraction of the total H$\alpha$ flux in
the offset line components as a function of the same. The Pearson
correlation coefficient is  given in the two bottom panels.
\label{models}}
\end{figure}

Figure 8 shows that F(H$\alpha$) varies between 0.05 and 0.25, with fairly
strong dependence on stellar mass, but no dependence on mean stellar age.
In contrast, in the sample of high redshift galaxies and AGN studied
by F\"orster-Schreiber at al (2019), the flux in the broad components often
exceeded the flux in the narrow component by a factor of two at the highest
stellar masses and in the AGN. Our measurements of V80 appear to be divided
into two well-separated classes -- galaxies where the excess emission is offset
by less than 200 km/s from the systemic velocity and those where the excess
emission extends beyond 500 km/s.  The latter class are almost always found in
galaxies with stellar masses greater than $10^{10} M_{\odot}$ and the former
class are found uniformly across the entire stellar mass range. V80 also does
not depend on the mean stellar age of the galaxy. Figure 9 shows the same two
quantities as a function of the Balmer decrement H$\alpha$/H$\beta$ and the
extinction corrected H$\alpha$ equivalent width. F(H$\alpha$) shows a
significant correlation with the Balmer decrement and a weaker dependence on
EQW(H$\alpha$). There is a tendency for systems with large  V80  to have higher
Balmer decrements, but this is not as pronounced as the stellar mass dependence
shown in Figure 8.

\begin{figure}
\includegraphics[width=92mm]{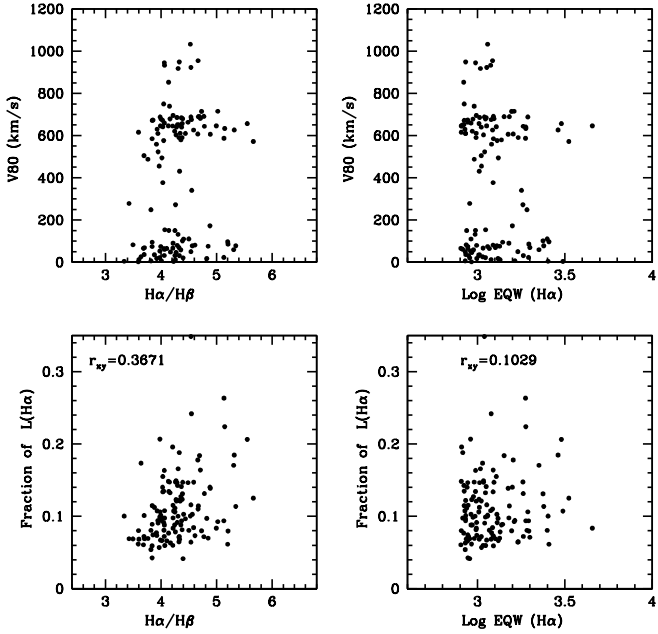}
\caption{The top panels show V80, the velocity separation enclosing 80\% of
the asymmetric flux, as a function the Balmer decrement and the logarithm
of the extinction corrected H$\alpha$ equivalent width of  
the galaxies in the H$\alpha$ excess sample. 
The bottom panels show F(H$\alpha$), the fraction of the total H$\alpha$ flux in
the offset line components as a function of the same.  The Pearson
correlation coefficient is  given in the two bottom panels.
\label{models}}
\end{figure}

Why is the distribution of the V80 parameter separated into two distinct peaks?
In Figure 10, we have divided H$\alpha$ excess sample with detectable H$\alpha$
line asymmetries into three separate classes. In the figure red points show
galaxies where the excess flux is on the red side of the H$\alpha$ line centrod,
blue points show galaxies where it is on the blue side and green points show
galaxies where excess emission is found on both sides of the line. We show the location of
these three classes in a number of  diagnostic plots.  Blue side
systems always have V80 less than 200 km/s. 
In red side systems,
the excess emission usually extends to velocities of 500 km/s or greater.
Red side systems also have higher Balmer decrement values than blue
side systems, particularly for lower ionization ([OIII]/H$\beta$ ratio) systems 
Interestingly, the average  F(H$\alpha$)  values for blue and red side
systems are the same ($\sim$ 0.1).

We also note that there appears to be a third, sparsely populated class of
galaxies with  V80$\sim$1000 km/s and where galaxies with both blue and red side
components are found. These galaxies, shown as green points, have
the highest [OIII]/H$\beta$ ratios  and a significant fraction are  found in the
region of the BPT diagram occupied by AGN and composite galaxies.

\begin{figure}
\includegraphics[width=92mm]{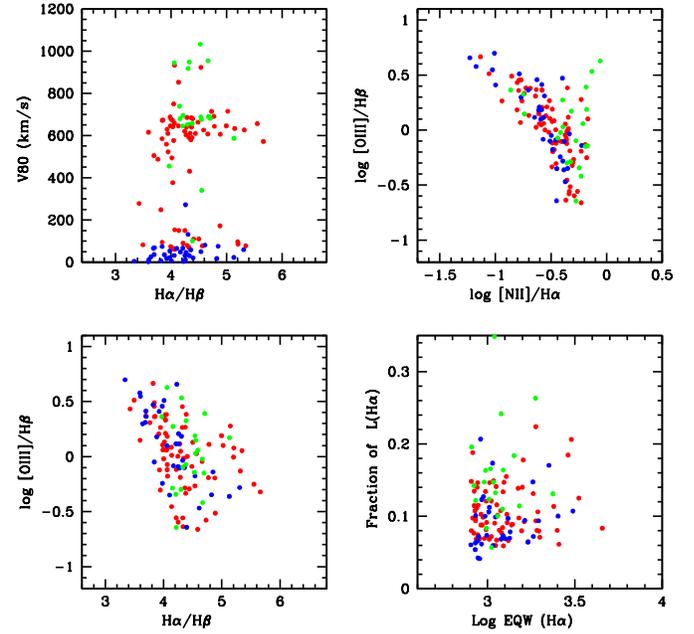}
\caption{The top left panel shows V80, the velocity separation enclosing 80\% of
the asymmetric flux as a function the Balmer decrement. 
The top right panel shows the [OIII]/H$\beta$ versus [NII]/H$\alpha$ BPT diagram.
The bottom left panel shows [OIII]/H$\beta$ versus the Balmer decrement.
the bottom right panel shows the F(H$\alpha$) as a function of 
the logarithm of the extinction corrected H$\alpha$ equivalent width.
\label{models}}
\end{figure}

We note that Concas et al (2019) studied the incidence of neutral and ionized
gas outflows in typical samples of galaxies in the local Universe and found that
extended high-velocity  H$\alpha$ emission was only found in AGN and not in the
normal star-forming galaxy population.  F\"orster-Schreiber et al (2019) studied
H$\alpha$ line profile properties in a sample of typical galaxies at redshifts
0.6-2.7 and found that extended H$\alpha$  emission was found in both
star-forming galaxies and AGN, but that there was a  strong dichotomy in the
velocity extent of the H$\alpha$ emission between star-forming galaxies and AGN.
Here we have selected a subset of the most strongly star-forming galaxies (as
traced by their H$\alpha$ emission) in the local Universe. 
The galaxies
plotted in green in Figure 10 with the highest values of F(H$\alpha$)
and V80 appear to be the closest analogues of
the AGN studied by Concas et al (2019) and  F\"orster-Schreiber et al (2019).
AGN are only apparent in a
very small minority of our H$\alpha$ excess  sample, but
insofar as they are found, they do seem to exhibit high velocity outflows
of ionized gas. 

It is then reasonable to speculate that the blue-shifted, low velocity H$\alpha$ components 
correspond to outflows generated by young stars and that broadened components
appear on the blue side of the line,  because the
emission on the red side of the line is preferentially absorbed by dust. What then are
the red-shifted, intermediate velocity (500-600 km/s) components?  One possibility
is that we are seeing inflowing rather than outflowing material that is triggering
the formation of the very massive star population. Further study of well-resolved 
nearby systems  using IFU data would  be very useful to investigate these issues in
more detail.

Finally, in Figure 11 we compare the relations between V80 and F(H$\alpha$) and
stellar mass $M_*$ for galaxies with detected line asymmetries in the H$\alpha$
excess (red) and control sample (black). As we previously noted, there are
many fewer such galaxies in the control sample. They appear to be
shifted to somewhat higher stellar masses, but have similar distributions of V80
and F(H$\alpha$) at fixed $M_*$.

\begin{figure}
\includegraphics[width=92mm]{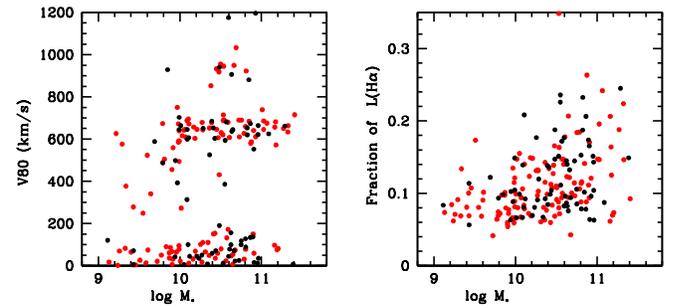}
\caption{The relations between V80 and F(H$\alpha$) and
stellar mass $M_*$ for galaxies with detected line asymmetries are plotted
for galaxies in the  H$\alpha$
excess (red) and control samples (black).
\label{models}}
\end{figure}

\subsection{Wolf Rayet features} Our procedure for selecting candidate Wolf
Rayet galaxies identified 62 blue bump and 69 red bump objects in the H$\alpha$
excess sample, and 54 blue bump and 52 red bump objects in the control sample.
The distribution of $S/N$ values for the detections is shown in Figure 12. The
average $S/N$ of the blue bump detections in the control sample is higher than
in the H$\alpha$ excess sample. In section 4, we will explain the cause of this
using stacked spectra. The $S/N$ values for the red bump detections in the
H$\alpha$ excess sample is slightly higher than in the control sample, but the
difference is quite small.

\begin{figure}
\includegraphics[width=92mm]{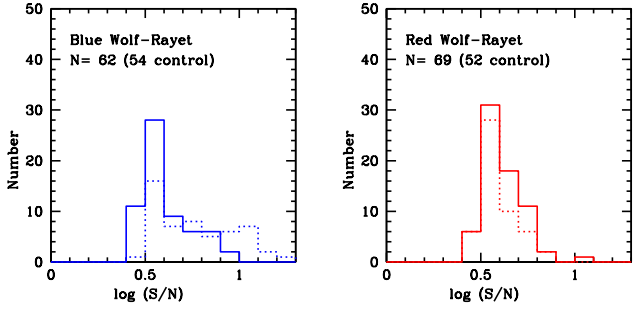}
\caption{Left: The solid and dashed blue histograms show the $S/N$ distribution
of the blue bump detections for the H$\alpha$ excess and control samples.
Right: The solid and dashed red histograms show the $S/N$ distribution
of the red bump detections for the H$\alpha$ excess and control samples.
\label{models}}
\end{figure}

In summary, the H$\alpha$ excess sample contains 15-20\% more galaxies with suspected Wolf
Rayet features compared to the control sample. This is considerably smaller than
the factor 2-3 boost that was quoted in Kauffmann (2021). This likely indicates
that Wolf Rayet signatures  become more strongly diluted if the spectra include
additional light from the outer galaxy. As we will show in the next section,
physical information about the stellar populations of blue bump Wolf Rayet
galaxies can still be extracted from the stacked spectra of galaxies at $z \sim
0.2$.

In the top panels of Figure 13, we plot mean stellar age as a function of stellar
mass (left) and the Balmer decrement (right) for all galaxies from the H$\alpha$
excess sample (small black points), for blue bump Wolf Rayet candidates with
$S/N >4$ (blue points) and for red bump Wolf Rayet candidates with $S/N >4$ (red
points). In the bottom panels we plot the metallity indicator O3N2 as a function
of the same two quantities.  The O3N2 index depends on two (strong) emission
line ratios and was first introduced and defined by Alloin et al. (1979) as
O3N2=log([OIII]$\lambda$5007/H$\beta$-log([NII$\lambda$6583/H$\alpha$).  Figure
13 shows that the blue and red bump Wolf Rayet candidates separate fairly
strongly by stellar mass and metallicity, with the blue bump objects located
mainly in galaxies with stellar masses less than $10^{10} M_{\odot}$ and the red
bump objects distributed fairly uniformly across the whole mass range from
$10^9-10^{11} M_{\odot}$. 

We will show in the next section that
our blue bump detections are confirmed as true Wolf Rayet galaxies in stacked spectra
by the detection of a clearly broadened HeII$\lambda$4686 line, but the same
is not true for our red bump detections. We thus caution the reader against
reading too much into these findings at this stage. 
We also note that the Wolf Rayet candidates have somewhat younger mean stellar age estimates
at fixed stellar mass than the underlying sample, but the age shift is rather
weak and is only apparent for more massive galaxies. There is no offset in
mean stellar age at fixed Balmer decrement H$\alpha$/H$\beta$. The weak shifts
in mean stellar age may indicate that the youngest stars are only  a very small
fraction of the total stellar mass probed by the fibre. Alternatively,
the light from many of the youngest stars may be heavily attenuated by dust.
We will discuss this in more detail in the final section of the paper.

\begin{figure}
\includegraphics[width=92mm]{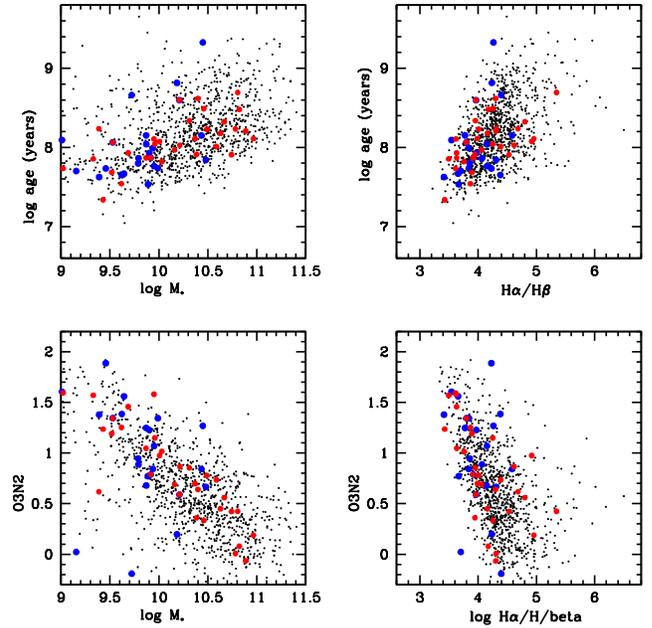}
\caption{Top panels: mean stellar age is plotted as a function of stellar
mass (left) and the Balmer decrement (right) for all galaxies from the H$\alpha$
excess sample (small black points), for blue bump Wolf Rayet candidates with
$S/N >4$ (blue points) and for red bump Wolf Rayet candidates with $S/N >4$ (red
points). Bottom panels: the metallity indicator O3N2 is plotted as a function
of the same two quantities.
\label{models}}
\end{figure}

\subsection{Radio detections} We recover 208 VLA FIRST survey cross-matches in the
H$\alpha$ excess sample, compared to 134 in the control sample, i.e. a 55\%
higher detection rate.  Figure 14 shows histograms of the stellar masses and
mean stellar ages of the galaxies in the radio-loud subsamples compared to their
parent samples.  In the left panels, we compare the radio-loud subsample
extracted from the H$\alpha$ excess sample (solid histograms) with the parent
H$\alpha$ excess sample (dotted histograms).  The radio-loud samples are shifted
to higher stellar masses and older mean stellar ages. In the middle panels, we
compare the radio-loud subsample extracted from the control samples and find a
similar shift to larger stellar masses and older ages.  In the right panels, we
compare the radio loud subsamples from the H$\alpha$ excess and control samples
with each other, finding a shift towards younger stellar populations for
radio-loud galaxies from the H$\alpha$ excess sample in comparison to the
radio-loud control galaxies. These controlled  comparisons yield evidence that the
higher level of radio activity        
amongst the H$\alpha$ excess galaxies may be tied to the presence of young stars.

\begin{figure}
\includegraphics[width=92mm]{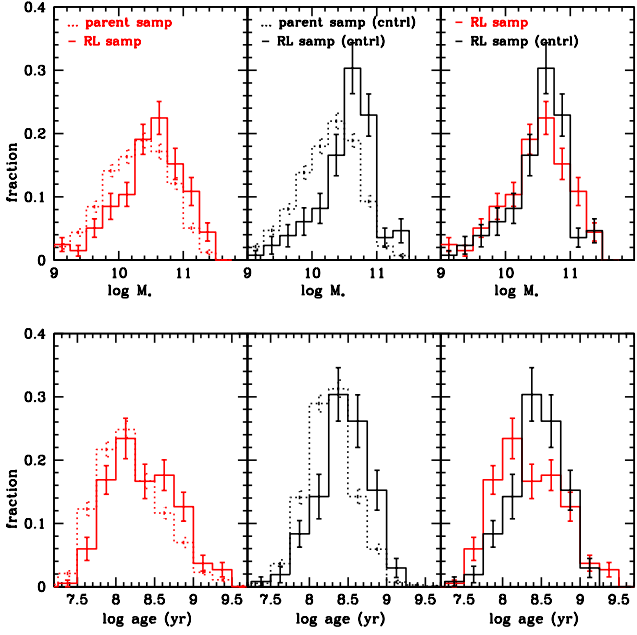}
\caption{Histograms of the stellar masses and
mean stellar ages of the galaxies in the radio-loud subsamples compared to their
parent samples. Left panels: the radio-loud subsample
extracted from the H$\alpha$ excess sample (solid histograms) and  the parent
H$\alpha$ excess sample (dotted histograms). Middle panels:
the radio-loud subsample extracted from the control samples (solid)
and the full control sample (dotted). Right panels: the radio 
loud subsamples from the H$\alpha$ excess (red solid)  and control samples (black solid).
Error bars on the histograms are computed via a standard bootstrap
resampling technique.
\label{models}}
\end{figure}

Figure 15 explores relations between the radio luminosities of the galaxies
with VLA FIRST detections in the H$\alpha$ excess sample (red points) and in the
control sample  (black points) with different host galaxy properties.
The top left panel shows that there is a strong
correlation between radio power P and stellar mass in both samples, with the
control sample galaxies shifted to higher stellar masses as indicated in Figure
14.  In the next three panels, we scale out the stellar mass dependence and plot
the ratio of radio power to stellar mass as a function of H$\alpha$ equivalent
width, Balmer decrement and mean stellar age. The quantity $\log P/M_*$ only
shows a fairly strong residual dependence with mean stellar age.  $P/M_*$ is on
average a factor of 10 larger for galaxies with mean stellar ages of a few$
\times 10^7$ years than it is for galaxies with stellar ages of $\sim 10^9$
years. This again provides evidence that the radio emission is linked
to the presence of  young stars.

\begin{figure}
\includegraphics[width=92mm]{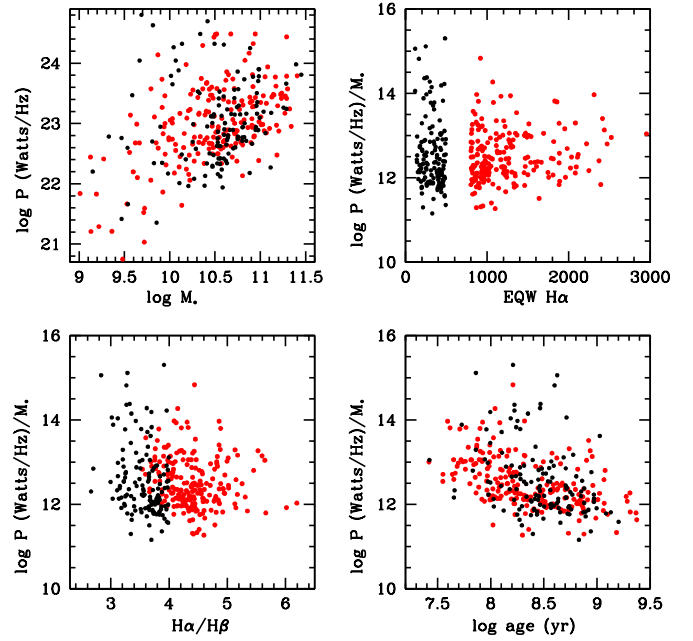}
\caption{Top left: correlation between radio power P and stellar mass.
VLA FIRST detections in the H$\alpha$ excess sample are plotted as red
points and in the control sample as black points.
Top right: the ratio of radio power to stellar mass ( $\log P/M_*$)  
as a function of H$\alpha$ equivalent width. Bottom left: $\log P/M_*$
as a function of Balmer decrement. Bottom right: $\log P/M_*$
as function of mean stellar age.
\label{models}}
\end{figure}

FIRST cut-out images of the most radio-luminous galaxies in the sample 
to see if there is any evidence for  AGN-powered jets in these galaxies and we find that the
majority are unresolved point sources. There are a handful of galaxies where
more than one point source is detected and these are usually merging/interacting
systems.  The brightest source in the sample with $\log (P/{\rm Watts Hz}^{-1}) > 25$ does
exhibit a classical jet-lobe morphology. Since it is the only one in the sample
and the most extreme object, we remove it when carrying out the spectral
stacking analysis described in the next section.

\section {A search for additional physical information using stacked spectra}

We now utilize stacked spectra as a way of gaining further insight into the main
physical processes that could be at work in the H$\alpha$ excess sample. As
outlined in section 1, it is of particular interest to investigate whether
galaxies with evidence for unusual populations of young, massive stars may also
be sites for the formation of intermediate mass black holes, or whether
efficient accretion onto existing black holes may be occuring in such systems.

We have shown that the majority of galaxies fall well within the star-forming
locus in the canonical [OIII]/H$\beta$ versus [NII]/H$\alpha$ BPT diagram, so
any AGN contribution to the [OIII] line is heavily swamped by emission from the
HII regions in these galaxies. Higher ionization lines, such as [NeV]$\lambda$3345
and [NeV]$\lambda$3425 provide possible AGN diagnostics where ionization from
young stars is likely to be much less important. It has been shown that highly
obscured, Compton thick (column density $N_H > 10^{24}-10^{25}$ sm$^{-2}$) AGN
hosted in massive star-forming galaxies sometimes show strong [NeV] emission
(Gilli et al 2010; Lanzuisi et al 2018).  We note that the detection of [NeV] is
not, in and of itself, an existence proof of an accreting black hole. Izotov,
Thuan \& Privon (2012) have identified [NeV] emission lines in a number of blue
compact dwarf galaxies and favour an explanation where the emission arises from
fast, radiative shocks associated with the starburst itself.

In this analysis, we have split the full H$\alpha$ excess sample into
sub-samples according to properties such as 1) radio luminosity, 2) presence or
absence of H$\alpha$ line asymmetries extending to high velocities, and 3)
presence or absence of Wolf Rayet signatures. We create stacked spectra for each
of these subsamples and investigate key spectral regions for systematic trends.
For example, if the strength of the [NeV] line were found to scale with the radio
luminosity of the system, this might provide evidence that shocks
induced by a relativistically expanding jet were responsible
for both the radio emission and the excitation of the hard ionizing radiation.

\subsection {Spectra stacked by radio luminosity}

Figure 16 shows spectra stacked according their radio properties -- we plot
spectra for H$\alpha$ excess sample galaxies with no radio detection in black,
for all galaxies with a VLA FIRST radio detection in green, for galaxies with
$22.5 < \log (P/{\rm Watts Hz}^{-1}) < 23.5$  in blue, and for galaxies with $23.5 <
\log P/{\rm Watts Hz}^{-1}) < 24.5$  in red. The top right panel shows the spectral
range covering the [NeV]$\lambda$3345 and [NeV]$\lambda$3425 emission lines.  There
is no detectable [NeV] emission for the sample without radio detections.  Only for
the two highest radio luminosity bins are  there  clear detections of 
[NeV]$\lambda$3425. The line is broadened in comparsion to the [OIII] lines shown
in the bottom left panel of the figure. The [NeV]$\lambda$3425 line exhibits a
hint of a double-peaked profile, which has been proposed as a signature of
jet-interstellar medium interactions in galaxies (e.g. Rubinur \& Kharb 2019,
Kharb et al 2021).

\begin{figure*}
\includegraphics[width=130mm]{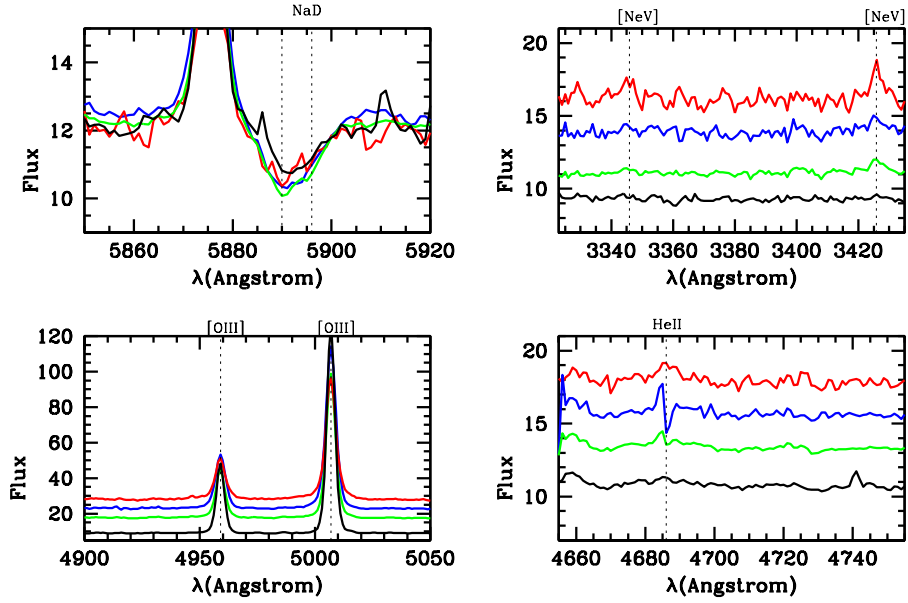}
\caption{Spectra stacked according their radio properties.
The stacked spectrum for H$\alpha$ excess sample galaxies with no radio 
detection is plotted in black, for all galaxies with a VLA FIRST radio detection in green, for galaxies with
$22.5 < \log P/{\rm Watts Hz}^{-1}) < 23.5$ in blue, and for galaxies with $23.5 <
\log (P/{\rm Watts Hz}^{-1}) < 24.5$  in red.
The top left panel shows the  region of the spectrum
around the Na I $\lambda\lambda$5890, 5896 (Na D) aborption line doublet 
(The position of the doublet is marked using dotted lines.)
 The top right panel shows the spectral
range covering the [NeV]$\lambda$3345 and [NeV]$\lambda$3425 emission lines.
The bottom left panel shows the spectral
range covering the [OIII]$\lambda$4959 and [OIII]$\lambda$5007 emission lines.
The bottom right panel  shows the spectral region around
HeII$\lambda$4686.
\label{models}}
\end{figure*}

The bottom right panel of Figure 16 shows the spectral region around
HeII$\lambda$4686, another high ionization emission line. The stacked spectrum
of the strongest radio sources again shows a broadened, double-peaked line.
Interestingly, HeII$\lambda$4686 in the stacked spectrum of the intermediate
luminosity sources with $22.5 < \log(P/{\rm Watts Hz}^{-1}) < 23.5$ exhibits a clear
inverse P-Cygni profile, with one component in emission bluewards of the
expected line centre, and another component in absorption on the red ride of the
line. Inverse P-Cygni profiles are believed to be a diagnostic of infalling gas.
They are a characteristic feature of molecular line observations of protostars.
The same line profile shape is still visible for the stack of the full sample of
radio-loud galaxies plotted in green, but the emission and absorption components
are much weaker. The stacked spectrum of galaxies without radio detections does
not exhibit HeII emission.

The two right panels of Figure 16 demonstrate that very high ionization emission
lines are strongest for the most luminous radio sources in the H$\alpha$ excess
sample.  Comparison of the [OIII]$\lambda$4959 and [OIII]$\lambda$5007 emission
lines for stacked spectra with different radio luminosities show that the
opposite is true. These lines get progressively {\em weaker} at high radio
luminosities. The [OIII] lines are also very regular with no clear broadened
components or asymmetries. This suggests that these lines trace very different
gas components. Unfortunately there is no information from single fibre spectra
about the spatial scale of the emission, but one might hypothesize from the
regularity of the line  profiles that the [OIII]-emitting gas is in virial
equilibrium with the stars in the galaxy, whereas the higher ionization lines
come from more localized sources.

Why are lines traced by the global galaxy gas component weaker in the radio-loud
systems? The top left panel of Figure 16 shows the  region of the spectrum
around the Na I $\lambda\lambda$5890, 5896 (Na D) sborption line doublet. This
absorption line has been used as a probe of the kinematics of cool, neutral gas
in galaxies (Chen et al 2010; Concas et al 2018). In normal star-forming
galaxies, the lines can be separated into two components: a quiescent disk-like
component at the galaxy systemic velocity and a blue-shifted outflow
component, which becomes stronger in galaxies with higher star formation surface
densities and dust content.  The galaxies in the H$\alpha$ excess sample are
found to have rather irregular morphologies on average, so it is perhaps not
surprising that the NaD doublets in the stacked spectra appear more blended that
in previous studies of normal  star-forming galaxies.  Nevertheless. Figure 16
shows a clear bluewards shift of the absorption line profile for the radio-loud
galaxies compared to the galaxies with no radio detections.  This is an
indication of more neutral outflowing gas in the radio-loud sample.  However,
there is no clear correlation of line shift  with radio luminosity -- all the
radio-loud subsamples have approximately the same NaD absorption line profiles.
This suggests that the main energy source for the outflows is not the same as
for the radio emission, and is most likely star formation.

\subsection {Spectra stacked according to the presence of high-velocity
H$\alpha$ line components}

In this section, we examine stacked spectra of two subsets of the H$\alpha$
excess sample: 1) those with no detectable H$\alpha$ line profile asymmetries,
2)those with ``fast'' outflows with $V80 > 500$ km/s. In Figure 17, the
stacked spectrum of subsample (1) is plotted in blue and that of subsample (2)
in red over the same wavelength ranges as in Figure 16.

Figure 17 shows that high ionization lines are weak in both spectral stacks and
show no differences according to whether or not high-velocity H$\alpha$ line
components are detected. The [OIII] line is somewhat weaker in the stack with
high velocity H$\alpha$ components and the NaD absorption feature shows a
stronger blueshift. These trends are consistent with the hypothesis that the
high-velocity H$\alpha$ components trace outflowing gas from the galaxy.

\begin{figure*}
\includegraphics[width=130mm]{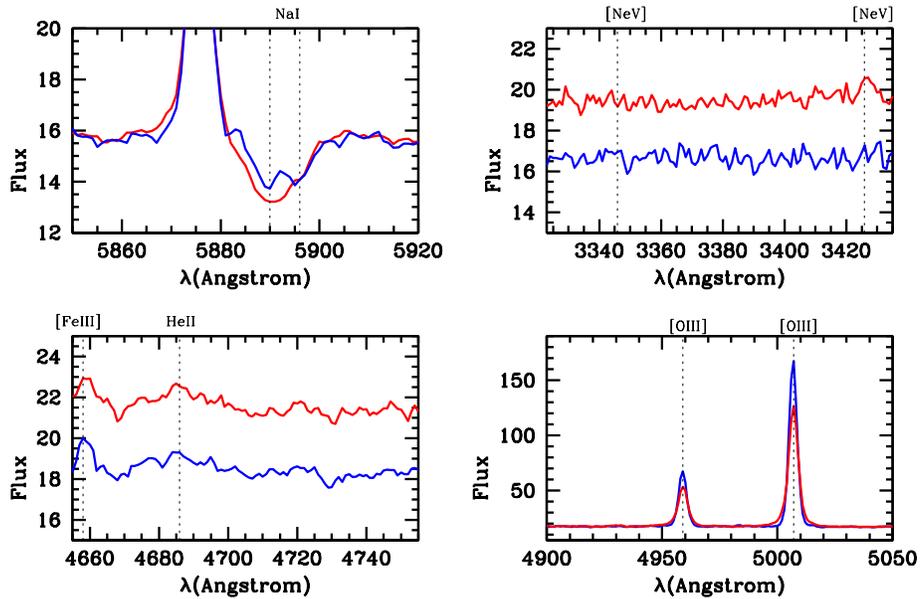}
\caption{As in Figure 16, but for spectra stacked according to whether or not
aysymmetric H$\alpha$ line components with  $V80 > 500$ km/s are found.
The stacked spectrum for H$\alpha$ excess sample galaxies with high velocity   
H$\alpha$ components is plotted in red, and for H$\alpha$ excess sample galaxies
with no detectable line asymmetries in blue. 
\label{models}}
\end{figure*}

\subsection {Spectra stacked according to the blue and red bump detections}
Figure 18 compares the stacked spectra of galaxies from the H$\alpha$ excess and
control samples with blue bump detections with $S/N >4$. The procedures for
measuring the blue and red bumps have been outlined in Section 2.2.  
The HeII$\lambda 4686$ emission line is strong in both the H$\alpha$ excess  and the
control sample stacked spectra.  In the H$\alpha$ excess stack,  HeII$\lambda$
4686 line is broader than the higher ionization [FeIII]$\lambda$4658 line 
(FWHM$\sim$ 15 \AA for  HeII compared to 3 \AA\ for [FeIII]). This is suggestive of a late 
WN population (Crowther \& Walborn 2011).  In the
control sample stack, the HeII $\lambda$4686 line is similar in width to
H$\beta$, suggesting a nebular origin. HeI is stronger in the  H$\alpha$ excess
stack and the high ionization [ArIV] lines are stronger in the control sample
stack.

\begin{figure*}
\includegraphics[width=130mm]{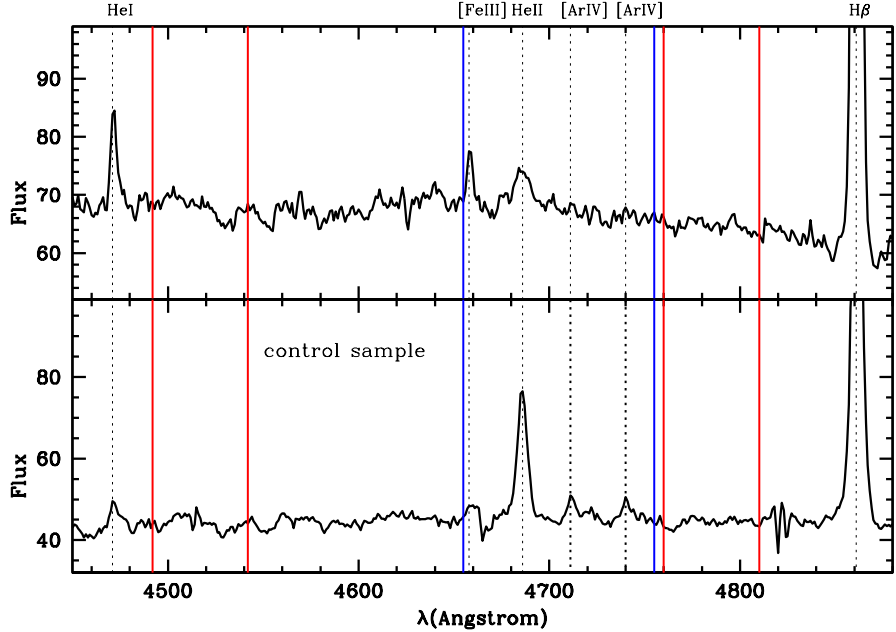}
\caption{Stacked spectra of galaxies from the H$\alpha$ excess (top) and
control samples (bottom) with blue bump detections with $S/N >4$
over the spectral region covering the blue bump.
The central bandpass is delineated by blue lines and the
two continuum bandpasses by red lines.
\label{models}}
\end{figure*}

Figure 19 shows the same comparison for galaxies with red bump detections in
the two samples. The stack from the H$\alpha$ excess sample is characterized
by strong interstellar absorption features. The strongest of these is the NaD
doublet marked in the figure. In many starburst galaxies, the only narrow
emission line that is usually detected in red bump identifications is
[NII]$\lambda$5755 (see Figure 2 of  L\'opez-S\'anchez \& Esteban (2018)). This
line is clearly visible in our H$\alpha$ excess stack, but not in the control
sample stack.

\begin{figure*}
\includegraphics[width=130mm]{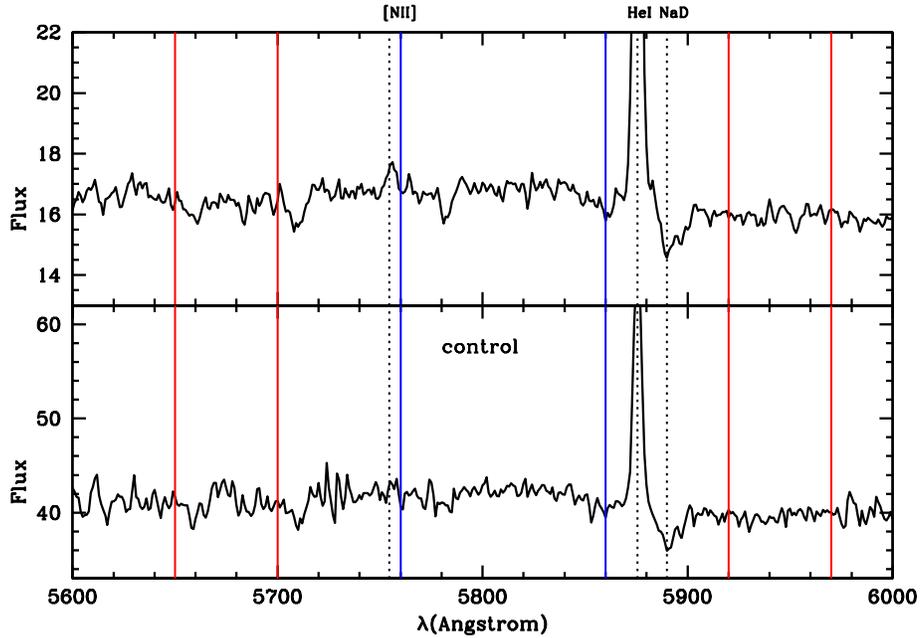}
\caption{Stacked spectra of galaxies from the H$\alpha$ excess (top) and
control samples (bottom) with red bump detections with $S/N >4$
over the spectral region covering the red bump.
The central bandpass is delineated by blue lines and the
two continuum bandpasses by red lines.
\label{models}}
\end{figure*}

In Figure 20, we examine the spectral region covering [NeV]$\lambda$3345 and
[NeV]$\lambda$3425 (left panels) and [NeIII]$\lambda$3869 and [NeIII]$\lambda$3967
(right panels) for the 4 sets of stacked spectra shown in Figures 18 and 19. The
[NeV] lines are not detected in either of the H$\alpha$ excess stacks. Strong [NeV]
lines are detected only in the control sample stack. The lines are similar in
width and shape to the  HeII$\lambda$ 4686 line shown in the bottom panel of
Figure 18, so it is likely that the source of ionization is the same for both.
Tha ratio HeII$\lambda 4686$/H$\beta$ in this stack is 0.34, which would
classify them as AGN in the HeII emission line diagnostic diagram introduced by
Shirazi \& Brinchmann (2012). These authors showed that HeII-strong AGN were
found mainly in  star-forming galaxies on the blue cloud and on the main
sequence where ionization from star formation is most likely to mask AGN
emission in the BPT lines. In follow-up work, B\"ar et al (2017) cross-matched a
sample of 234 He II-only AGN candidates with the Chandra Source Catalog (Evans
et al. 2010).  Among 12 objects with X-ray detections,  five objects were
confirmed as AGN based on their X-ray luminosity and power-law nature; of the
remaining seven objects, six objects had X-ray luminosity upper limits
consistent with being AGN.

\begin{figure*}
\includegraphics[width=130mm]{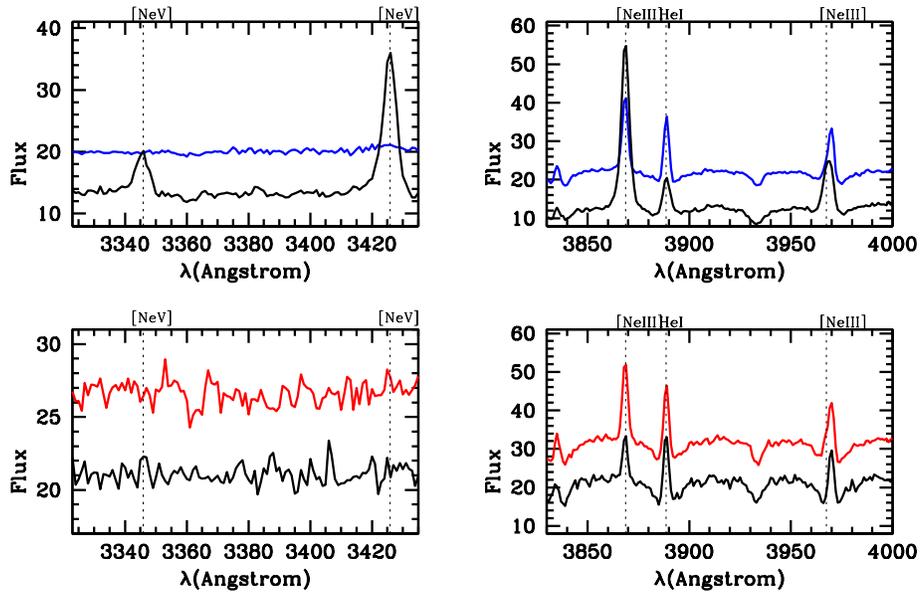}
\caption{The spectral region covering [NeV]$\lambda$3345 and
[NeV]$\lambda$3425 (left panels) and [NeIII]$\lambda$3869 and [NeIII]$\lambda$3967
(right panels) for the 4 sets of stacked spectra shown in Figures 18 and 19.
The blue bump and control stacks are shown in the top panels, plotted
in blue and black.
The red bump and control stacks are shown in the top panels, plotted
in red and black.
\label{models}}
\end{figure*}

We have postulated that our control sample galaxies represent a later stage of a
stong star-forming episode compared to the H$\alpha$ excess galaxies. In the
H$\alpha$ excess sample, the HeII$\lambda 4686$ emission is found to be produced
by Wolf Rayet stars and in the control sample, the HeII$\lambda 4686$ is likely
produced by an accreting black hole. It is thus very tempting to speculate that
there is an evolutionary sequence from an environment that is very rich in
massive stars to an environment that contains one or more black holes in
formation.

\section{Summary of findings and future perspectives} We have selected two
samples from the SDSS main galaxy sample on the basis of their
extinction-corrected H$\alpha$ equivalent widths.  The first sample is selected
to have EQW(H$\alpha$/\AA)$>800$ and is called the H$\alpha$ excess sample, because
EQW(H$\alpha$) is too high to be explained by ionization by a young stellar
population with a normal initial mass function (IMF). We create a control sample
where galaxies are matched to each H$\alpha$ excess galaxy in stellar mass,
redshift and 4000 \AA\ break strength, but where the H$\alpha$ EQW is in the
range 80-300 \AA, typical of strongly star-forming galaxies with a normal IMF.

We carry out a systematic comparison of the two samples, with the following main
findings: \begin{itemize} 

\item The H$\alpha$ excess galaxies have median
stellar mass of $2 \times 10^{10} M_{\odot}$.  They have younger mean stellar
ages and more dust than the control sample galaxies.  Almost all galaxies in
both samples lie within the star-forming locus in the [OIII]/H$\beta$ versus
[NII]/H$\alpha$ BPT diagram.  

\item H$\alpha$ excess galaxies are twice as
likely to exhibit H$\alpha$ line profile asymmetries compared to control sample
galaxies.  

\item The fraction of the total H$\alpha$ flux in the asymmetric
components ranges between 0.05 and 0.25 and is larger for more massive galaxies
and galaxies with higher dust extinction. 

\item Two distinct types of H$\alpha$
line profile asymmetries are identified: a) Blue-shifted  H$\alpha$ components
that extend over velocity separations of less than 200 km/s from
the systemic redshift b) Red-shifted H$\alpha$ components
that extend to velocities of 500 km/s or greater.  The galaxies with the  highest
velocity H$\alpha$ components ($>1000$ km/s) have BPT line ratios indicative of
AGN ionization. 

\item H$\alpha$ excess galaxies are 1.55
times as likely to have radio detections in the VLA FIRST catalogue compared to
control sample galaxies. In the H$\alpha$ excess sample, radio luminosity per 
unit stellar mass is a factor of 10 larger for galaxies with stellar ages
$\sim 10^{7.5}$ yr compared to galaxies with stellr ages $\sim 10^{9}$ yr.    

\item The stacked
spectra of H$\alpha$ excess galaxies with the highest radio luminosities exhibit
high ionization [NeV]$\lambda$3345, [NeV]$\lambda$3425 and HeII$\lambda$4686
emission.  The line shapes of the high ionization lines are complex:
they are sometimes broadened and they sometimes exhibit inverse P-cygni profiles indicative
of gas inflow.  

\item We search for emission from very young Wolf Rayet stars by
looking for excess emission over the wavelength ranges 4665-4755 \AA\ and
5760-5860 \AA. Similar numbers of candidates are found for both the H$\alpha$
excess and control samples. The stacked spectrum of WR candidates from
the  H$\alpha$ excess sample reveals a broadened HeII$\lambda$4686 line
characteristic of WN star emission and no high-ionization [NeV] lines.
In contrast, the stacked spectrum of WR candidates from
the control sample reveals a narrow HeII$\lambda$4686 line and strong
[NeV]$\lambda$3345 and [NeV]$\lambda$3425 emission characteristic of AGN.

\end{itemize}

In summary, we have utilized a  diverse range of indicators to probe the H$\alpha$ excess sample
for hidden populations of accreting black holes. We find a strong correlation between
radio luminosity per unit stellar mass and mean stellar age such that 
log P/$M_*$ increases for the systems with the youngest stellar populations.  We also find that
the [NeV] emission line strength correlates with radio luminosity in the  H$\alpha$
excess sample. [NeV] emission is not present in the very youngest radio-quiet H$\alpha$ excess galaxies with 
detectable Wolf-Rayet features. Although correlations of the kind presented in this paper
suggest a causal connection between star formation, black hole formation/accretion and
the generation of radio jets,  
a campaign to spatially map the emission   
from  representative samples of such objects is required to establish a  true picture
of what is happening within these very dense star-forming regions.

Our study has also revealed the promise of spectral stacking as a way of 
pulling out the weaker line emission associated with short-lived, high mass 
stars. Synthetic spectra of the most massive stars  calculated
from  model atmospheres which account for non-LTE, spherical expansion and
and metal line blanketing (Hamman \& Gr\"afener 2004) can be incorporated into 
stellar population synthesis codes. The HR-pyPopStar model (Mill\'an-Irgoyen et al 2021)
provides a complete set of
high resolution spectral energy distributions of single stellar populations.
The model incorporates  high wavelength-resolution theoretical atmosphere libraries
for main sequence, post-AGB/planetary nebulae and Wolf-Rayet stars and is an update of the
models presented in Moll\'a, Garc\'ia-Vargas \&  Bressan  (2009).
Figure 21 shows HR-pyPopStar SSPs plotted over the wavelength range that includes
the main blue Wolf Rayet features. The spectra have been smoothed to 2 \AA
resolution to be more directly comparable to our SDSS spectra. Results are shown
for solar metallicity models and a Chabrier (2003) IMF at six different times 
from $10^6-10^7$ years. The development of a broad feature around the HeII $\lambda$4686
line commences at an age of 3$\times 10^6$ years and is over by an age of
5$\times 10^6$, i.e. it is a very short phase in the evolution of a simple 
stellar population. The [FeIII]$\lambda$4658  line is present in emission 
until a time of  3$\times 10^6$ yr. The blue spectrum plotted in the left middle
panel of Figure 21 shows the  HR-pyPopStar SSP for a 0.4 solar model
for comparison. [FeIII]$\lambda$4658 emission is not present at this lower
metallicity. In Figure 22, we overplot  HR-pyPopStar solar metallicity SSPs
at time $\log (t/{\rm yr}) = 6.48$ (red solid lines) and  $\log (t/{\rm yr}) = 6.54$ (red dotted lines)  
on the stacked spectrum of  H$\alpha$ excess sample galaxies with blue bump detections
(black solid lines). Both SSPs fit the  [FeIII]$\lambda$4658 line.
The width of the broad  HeII $\lambda$4686 is in good agreement with the data,
but the amplitude of the feature is too low to match the observations.

\begin{figure}
\includegraphics[width=92mm]{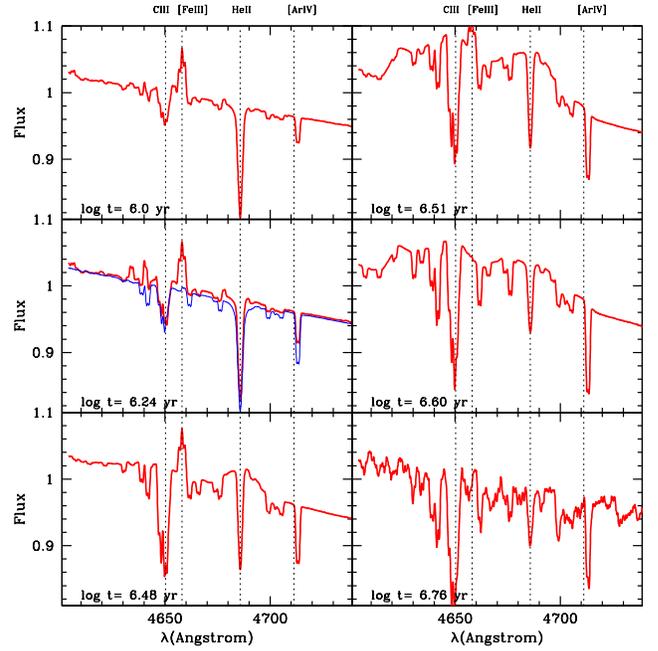}
\caption{HR-pyPopStar SSPs plotted over the wavelength range that includes
the main blue Wolf Rayet features. The spectra have been smoothed to 2 \AA
resolution to be more directly comparable to our SDSS spectra. Results are shown
for solar metallicity models and a Chabrier (2003) IMF at six different times
from $10^6-10^7$ years. The blue spectrum plotted in the left middle
panel of Figure 21 shows the  HR-pyPopStar SSP for a 0.4 solar model
for comparison. 
\label{models}}
\end{figure}

Past work has also found that  SSP models underpredict
blue optical Wolf Rayet bumps. Sidoli et al (2006) analyzed the      
spectrum of the  giant HII region Tol89 in NGC5398, inferring the O star content from
the stellar continuum and using STARBURST99 models to predict optical WR features using the grids from
Smith et al. (2002). They found that the models underpredict the WR features and 
attributed this failure to the neglect of rotational mixing in evolutionary models.
Evolutionary mixing increases the lifetimes of WR phases as a result of the 
increased duration of the H-rich phase, and lowers
the initial mass limit for the formation of WR stars. This compromises the
use of the WR features as diagnostics of other physical parameters such as
stellar IMF or recent star formation history.

\begin{figure}
\includegraphics[width=92mm]{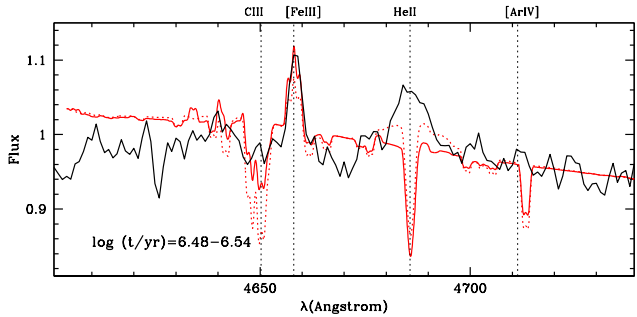}
\caption{ HR-pyPopStar solar metallicity SSPs
at times $\log (t/{\rm yr})= 6.48$ (red solid lines) and  $\log (t/{\rm yr})= 6.54$ (red dotted lines)
are overplotted on the stacked spectrum from the H$\alpha$ excess sample with blue bump detections
(black solid lines).
\label{models}}
\end{figure}

It is possible that by constructing grids of stacked empirical spectra from
large galaxy samples drawn from IFU surveys such as MaNGA,
and by studying how Wolf Rayet and continuum  features scale with each other
in a variety of galactic environments, insights may be obtained that serve
to better constrain both stellar models and the possibly changing nature of young stellar
populations in dense stellar environments. This will be the subject of future work.

Finally we note that many of the galaxies in the H$\alpha$ excess sample are quite
highly reddened. This can be seen very clearly in Figure 22 from the negative offset in the
observed continuum compared to the model continuum on the blue side of the wavelength
window. Correcting for the effect of dust on spectral lines requires a model for
the spatial distribution  of the dust in HII regions. Simple models, e.g. 
Charlot \& Fall (2000), may break down in very extreme galactic environments and
follow-up near-infrared observations (e.g. Crowther  et al 2006, Rosslowe \& Crowther 2018)
can serve to better constrain the true content of Wolf-Rayet stars in galaxies.

\vspace{4mm}
\noindent
{\bf Acknowledgments}\\
\normalsize
GK thanks Selma de Mink for helpful discussions related to this work.

\vspace{3mm}
\noindent
{\bf Data Availability}\\
\normalsize
The data underlying this article are available in the article and in its online supplementary material.


\end{document}